\documentclass{iopart}

\usepackage[svgnames]{xcolor}
\usepackage[bookmarks=false,linkcolor=Navy,urlcolor=Navy,colorlinks,citecolor=Navy]{hyperref}
\usepackage{graphicx}
\usepackage[normalem]{ulem}
\newcommand{\figref}[2]{\hyperref[#1]{\ref{#1}(#2)}}
\newcommand{\eqref}[1]{{(\ref{#1})}}
\newcommand{\ket}[1]{|#1\rangle}

\begin{document}

\title{Majorana zero modes induced by superconducting phase bias}

\author{Omri Lesser}
\address{Department of Condensed Matter Physics, Weizmann Institute of Science, Rehovot, Israel 7610001}

\author{Yuval Oreg}
\address{Department of Condensed Matter Physics, Weizmann Institute of Science, Rehovot, Israel 7610001}

\begin{abstract}
Majorana zero modes in condensed matter systems have been the subject of much interest in recent years. Their non-Abelian exchange statistics, making them a unique state of matter, and their potential applications in topological quantum computation, earned them attention from both theorists and experimentalists. It is generally understood that in order to form Majorana zero modes in quasi-one-dimensional topological insulators, time-reversal symmetry must be broken. The straightforward mechanisms for doing so---applying magnetic fields or coupling to ferromagnets---turned out to have many unwanted side effects, such as degradation of superconductivity and the formation of sub-gap states, which is part of the reason Majorana zero modes have been eluding direct experimental detection for a long time. Here we review several proposal that rely on controlling the phase of the superconducting order parameter, either as the sole mechanism for time-reversal-symmetry breaking, or as an additional handy knob used to reduce the applied magnetic field. These proposals hold practical promise to improve Majorana formation, and they shed light on the physics underlying the formation of the topological superconducting state.
\end{abstract}

\tableofcontents

\ioptwocol

\section{Introduction}

Topology in condensed matter systems has been a major theme of research in the past decades~\cite{hasan_colloquium_2010, moore_birth_2010,qi_topological_2011,bernevig_topological_2013}. Starting with the discovery of the integer~\cite{klitzing_new_1980} and fractional~\cite{tsui_two-dimensional_1982} quantum Hall effects, which were explained later on using topological arguments~\cite{thouless_quantized_1982,niu_quantized_1985,girvin_quantum_1999}, the field has seen a massive growth in both theory and experiment. Noteworthy theoretical milestones are the discoveries of Chern insulator~\cite{haldane_model_1988}, which are analogs of the quantum Hall effect with zero net magnetic flux, and topological insulators~\cite{kane_quantum_2005,kane_$z_2$_2005}, which can be thought of as two copies of a Chern insulator such that time-reversal symmetry is preserved.

The fact that topology even appears in electronic systems has to do with the intrinsic geometry of quantum particles in a periodic potential: the Brillouin zone itself is a topologically non-trivial manifold, possibly with a large genus~\cite{volovik_universe_2009}. This inherent topology, and also the topology of the Hilbert space associated with certain Hamiltonians~\cite{avron_homotopy_1983,chiu_classification_2016}, give rise to a Berry phase~\cite{berry_quantal_1984}, which turned out to have implication on transport properties~\cite{thouless_quantized_1982,niu_quantized_1985}, and these are robust to various perturbations. Such robustness is referred to as ``topological protection": quantized properties remain intact as long as the perturbations do not exceed the energy gap of the system. Notice that we assumed the system indeed has a finite gap---this is an important property for all the systems we will discuss here.

Apart from its effects on transport properties, topology has implications with regard to the allowed quasiparticle excitations. In basic quantum mechanics we learn that there is a clear dichotomy between bosons and fermions, based on their exchange statistics. That is because when a particle moves around a particle identical to it, their two-body wavefunction must return to its original value. This, in turn, stems from the fact that the trajectory can be shrunk into a point, by taking it out of the plane.
However, in two dimensions that is not the case, and therefore other types of exchange statistics are possible~\cite{nayak_non-abelian_2008,stern_anyons_2008,lahtinen_short_2017}. These exotic quasiparticles are referred to as ``anyons", because they can have \emph{any} statistical phase (as opposed to just 0 or $\pi$, for bosons and fermions respectively)~\cite{nayak_non-abelian_2008,stern_anyons_2008,kitaev_anyons_2006}. Anyons, and in particular non-Abelian anyons, are predicted to have applications in topological quantum computation~\cite{lahtinen_short_2017}. Realizing them in an actual physical system is a challenging task.

One type of non Abelian anyon, perhaps the simplest one, is the Majorana zero mode (MZM)~\cite{bernevig_topological_2013,alicea_new_2012,karzig_universal_2016}. A Majorana fermion is essentially half of an ordinary fermion. This can be formulated using the language of second quantization: a fermion annihilation operator $c$ can be written as $c=\alpha+i\beta$, where $\alpha=\alpha^{\dagger}$, $\beta=\beta^{\dagger}$. The self-adjoint operators $\alpha$, $\beta$ are Majorana fermions. But there is more to it than just a formal basis transformation. In a seminal papers, Kitaev~\cite{kitaev_unpaired_2001} showed that \emph{unpaired} zero-energy Majorana end states appear at the edges of a one-dimensional (1D) $p$-wave superconducting (SC) chain. If one could realize such chains, it would be possible to move MZMs around each other (to \emph{braid} them), which would reveal their non-Abelian exchange statistics~\cite{alicea_non-abelian_2011}.

The phase of matter realized in the Kitaev chain is called a \emph{topological superconductor}~\cite{alicea_new_2012,leijnse_introduction_2012,sato_topological_2017}. One of the most important properties of this type of topological superconductor is that it breaks time-reversal symmetry; if that had not been the case, then any excitation would be doubled by virtue of the Kramers degeneracy, and we would not be able to get a single isolated MZM. In terms of topological classification~\cite{altland_nonstandard_1997,schnyder_classification_2008,kitaev_periodic_2009}, such topological superconductors belong to symmetry classes D or BDI, depending on whether an emergent time-reversal symmetry that squares to $+1$ is present. Ordinary SCs, however, are of the $s$-wave singlet type~\cite{tinkham_introduction_2004,gennes_superconductivity_1999}, which does not break time-reversal symmetry. Importantly, in this situation the fermion doubling cannot be evaded, and therefore isolated MZMs cannot be formed at the edges (only Majorana-Kramers pairs may appear). One therefore has to wisely engineer combinations of materials, including superconductivity and spin rotations that breaks time reversal symmetry  to get the desired behavior. In this review we will survey some of the prominent schemes that have been put forward for realizing MZMs, with a focus on semiconducting materials with strong spin-orbit coupling and various mechanisms of time-reversal-symmetry breaking. For a more thorough review, the reader is referred to Refs.~\cite{lutchyn_majorana_2018,leijnse_introduction_2012,sato_topological_2017}.

As we mentioned, time-reversal symmetry must be broken to realize isolated MZMs. Naturally, this has led the idea of applying an external magnetic field~\cite{lutchyn_majorana_2010,oreg_helical_2010,hell_two-dimensional_2017,pientka_topological_2017} or adding a ferromagnetic insulator to the system~\cite{pientka_topological_2013,nadj-perge_proposal_2013,nadj-perge_observation_2014,vaitiekenas_zero-bias_2020}. Approaches in which the proximitizing SC experiences a magnetic field (orbital or exchange) are, however, harmful to the proximity effect~\cite{sabonis_destructive_2020}. In such setups, magnetic impurity states may be formed inside the SC gap, and therefore the topological protection of the MZMs becomes weaker. In addition, these in-gap states hinder the reliable detection of MZMs. 
Moreover, applying a magnetic field to a SC may cause vortices, which hinder the MZM detection, and therefore these setups favor type-I SCs (like Al) which do not have vortices. This strongly limits the selection of host SCs, and usually prevents one from using type-II SCs like Nb whose SC gap is ten times that of Al.

The purpose of this review is to highlight an additional option for breaking time-reversal symmetry, which is biasing the SC phase. The fact that the SC order parameter is a complex number~\cite{tinkham_introduction_2004,gennes_superconductivity_1999} is a gift given to us by Nature: since we are already using SCs, we might as well utilize this extra degree of freedom. As we will see, controlling phase differences between SCs (or between segments of one SC) can facilitate the formation of a topological SC phase. Compared to more direct mechanisms of time-reversal-symmetry breaking, phase control is very gentle and can be quite harmless to the SC (see Sec.~\ref{sec:howto}). That is because the magnetic field is not applied to the sample itself, but rather its effect is focused to the region of interest by the flux loops. Therefore, various intriguing approaches taking advantage of this idea have appeared.

The rest of this review is organized as follows. In Sec.~\ref{sec:TISC}, we review the early proposals for combining topological insulators and ordinary phase-biased superconductors to create MZMs. In Sec.~\ref{sec:reducing} we describe proposals for using phase bias to \emph{reduce} the Zeeman field required to drive various semiconductor-superconductor hybrid systems into a topological phase. In Sec.~\ref{sec:phaseonly}, we review several suggestions for \emph{eliminating} the Zeeman field altogether, and realizing topological superconductivity using \emph{only} phase bias. We briefly touch upon the practical aspects of phase biasing a superconductor in Sec.~\ref{sec:howto}. We conclude and discuss avenues for future research in Sec.~\ref{sec:outlook}. 

\section{Topological insulator--superconductor hybrid system}\label{sec:TISC}
One of the first ideas on realizing topological superconductivity in a feasible system is due to Fu and Kane~\cite{fu_superconducting_2008}. It was a very appealing proposal, as it relied on the recently discovered topological insulators (TIs)~\cite{kane_$z_2$_2005,kane_quantum_2005,fu_topological_2007,moore_birth_2010} and standard $s$-wave superconductivity. Fu and Kane's approach is rooted on the somewhat intuitive notion that a combination of a topological insulator and a superconductor should give rise to a topological superconductor. As we shall see, that is true even without phase biasing the superconductor; the phase bias is only necessary to induce a Majorana bound state.

\begin{figure}
    \centering
    \includegraphics[width=\linewidth]{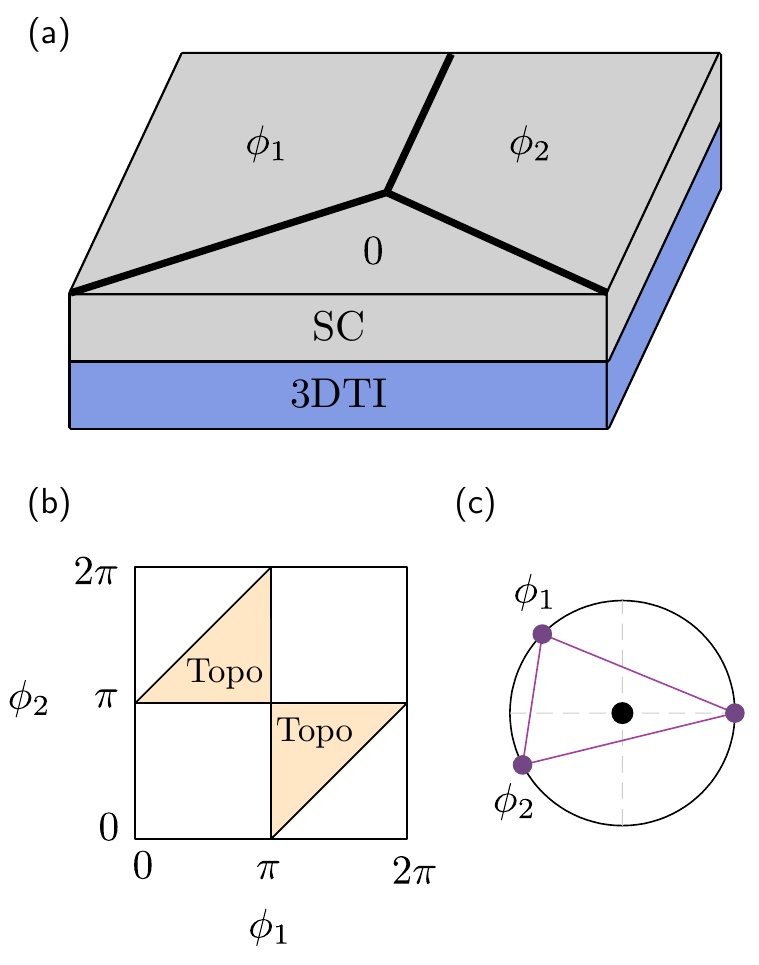}
    \caption{The Fu-Kane model~\cite{fu_superconducting_2008} for topological superconductivity in a trijunction. (a)~Model schematic: a superconductor (grey) with phase bias is proximity coupled to a 3D topological insulator (blue). (b)~Phase diagram as a function of the phases $\phi_1$ and $\phi_2$. The system becomes topological inside the triangles (orange) where the phases form a discrete vortex. (c)~The condition for a discrete vortex can be visualized geometrically as encircling of the origin by the triangle connecting $0$, $\phi_1$, $\phi_2$ when plotted as complex number on the unit circle~\cite{van_heck_single_2014}.}
    \label{fig:fu_kane}
\end{figure}

The analysis begins with the 2D surface states of a 3D topological insulator. These are described by the Dirac Hamiltonian $H_{\rm TI}=v\vec{\sigma}\cdot\vec{p}$, where $v$ is the Fermi velocity, $\vec{\sigma}$ is the vector of spin Pauli matrices, and $\vec{p}=-i\nabla$ is the 2D momentum. The TI is proximity coupled to an $s$-wave superconductor with a space-dependent phase, i.e., the pair potential is $\Delta\left(\vec{r}\right)=\Delta e^{i\phi\left(\vec{r}\right)}$, as shown in Fig.~\figref{fig:fu_kane}{a}. The full BdG Hamiltonian describing the system is then $H_{\rm FK}=\Psi^{\dagger}{\cal H}_{\rm FK}\Psi/2$ with
\begin{eqnarray}\label{eq:fu_kane_H}
    {\cal H}_{\rm FK} = &\left[-iv\left(\sigma_x \partial_x + \sigma_y \partial_y\right) -\mu \right]\tau_z \\
    &+ \Delta \left[\tau_x \cos\phi(\vec{r}) + \tau_y \sin\phi(\vec{r})\right]. \nonumber
\end{eqnarray}
Here, the $\tau$ Pauli matrices act on the particle-hole degree of freedom, and the Nambu spinor takes the form
\begin{equation}\label{eq:Nambu-spinor}
    \Psi=\left(\psi_{\uparrow}, \psi_{\downarrow}, \psi^{\dagger}_{\downarrow}, -\psi^{\dagger}_{\uparrow} \right)^{\rm T}.
\end{equation}

Like any BdG Hamiltonian, ${\cal H}_{\rm FK}$ respects particle-hole symmetry, as evident by $\left\{ {\cal H}_{\rm FK},\Xi \right\}=0$ where the particle-hole operator is $\Xi=i\tau_y \sigma_y {\cal K}$. 
Moreover and importantly, the TI itself does not break time-reversal symmetry: $H_{\rm TI}$ commutes with the time-reversal operator ${\cal T}=i\sigma_y{\cal K}$. The $s$-wave superconductor also does not break time-reversal symmetry, as long as it is not phase biased, i.e., if $\phi\left({\vec r}\right)$ is spatially uniform. In this time-reversal-symmetric case, the low-energy spectrum of the Fu-Kane Hamiltonian is similar to that of a spinless $p_x+ip_y$ superconductor~\cite{moore_nonabelions_1991} (they are not identical though: crucially, the $p_x+ip_y$ SC breaks TRS). Then, Fu and Kane followed an elegant line of thought: a vortex in a $p_x+ip_y$ SC is known to host a Majorana bound state at its core, so if the hybrid TI-SC system is phase biased such that a vortex is formed, it should also have a MBS.

At this point we stress an important issue that is often confusing with regard to the Fu-Kane model. As we mentioned in the previous paragraph, and as Fu and Kane showed by a simple canonical transformation, the analogy between the TI-SC system and the $p_x+ip_y$ SC is correct without phase biasing. In other words, the TI-SC system is inherently a topological superconductor. However, due to TRS it cannot have chiral edge modes. Moreover, in 2D, unlike in 1D, the formation of a topological superconducting state is not in one-to-one correspondence with the appearance of a Majorana zero mode. Therefore, the vortex is necessary in order to probe the topological nature of the TI-SC system, but it is not the one responsible for inducing it in the first place. It is the unique band structure of the TI that endows the hybrid system its topological nature.

A vortex may be introduced continuously, e.g. in polar coordinates $\phi\left(r,\theta\right)=n\theta$ (here the integer $n$ is the vorticity). Another possibility is inducing a \emph{discrete vortex}, as demonstrated in Fig.~\figref{fig:fu_kane}{a} for the case of a trijunction. When the phases $0$, $\phi_1$, $\phi_2$ wind, the system hosts a MBS, as shown in the phase diagram Fig.~\figref{fig:fu_kane}{b}. It has later been realized~\cite{van_heck_single_2014} that this winding is a fundamental and generic requirement for the existence of a zero-energy state in a phase-biased superconductor without magnetic fields. Visualizing the phases $0$, $\phi_1$, $\phi_2$ as complex numbers on the unit circle (see Fig.~\figref{fig:fu_kane}{c}), the existence of a discrete vortex is equivalent to the triangle connecting the phases encircling the origin. If the triangle does not encircle the origin, then the distance between the triangle and the origin sets a lower bound on the energy of bound states, prohibiting the existence of zero modes~\cite{van_heck_single_2014}.

Despite the apparent simplicity of the Fu-Kane model, its experimental realization turns out to be challenging. A main problem is that one needs a strong superconducting proximity effect at the surface states of the 3DTI, whereas in practice the bulk states tend to get better proximizited. The prominence of the bulk states, which are assumed in the theoretical model to be fully gapped and irrelevant, is adverse to the formation of a topological superconducting state. Another source of complication is the need of a very clean interface between the 3DTI and the SC, which is generally very hard to obtain. The cleanest 3DTIs are single crystal, but they are not as topologically robust as compounds (e.g. involving bismuth). Despite these major challenges, there have been some promising recent developments in inducing superconductivity in compound topological insulator~\cite{ghatak_anomalous_2018,bai_novel_2020}.

\section{Phase bias as a tool for reducing the Zeeman field}\label{sec:reducing}
In this section, we review ideas proposed to utilize the SC phase degree of freedom to lower the critical magnetic field needed for topological superconductivity. Although the magnetic field in these systems cannot go all the way to zero, its significant reduction makes them intriguing.

\subsection{Current-carrying nanowires}
Driving a constant current through at SC causes its phase to increase linearly~\cite{tinkham_introduction_2004}. This was exploited by Romito \etal~\cite{romito_manipulating_2012-1}, who studied the effect of supercurrent on the Majorana nanowires of Refs.~\cite{lutchyn_majorana_2010,oreg_helical_2010}. For context, we first review this model in the absence of current.

Consider a semiconducting spin-orbit-coupled wire, which is placed in proximity to an $s$-wave SC, with a magnetic field applied along the wire. The BdG Hamiltonian describing the system is~~\cite{lutchyn_majorana_2010,oreg_helical_2010} 
\begin{equation}\label{eq:majorana-nanowire}
    {\cal H}_{\rm NW} = \left( \frac{p^2}{2m}+up\sigma_x-\mu \right)\tau_z + E_{\rm Z}\sigma_z + \Delta\tau_x,
\end{equation}
where the Nambu spinor is the same as in Eq.~\eqref{eq:Nambu-spinor}, and the Pauli matrices $\sigma,\tau$ act in spin and particle-hole spaces, respectively. $m$ is the effective mass, $u$ is the SOC constant, $\mu$ is the chemical potential, $E_{\rm Z}$ is the Zeeman energy induced by the magnetic field, and $\Delta$ is the pair potential, which is chosen without loss of generality to be purely real. The normal-state spectrum ($\Delta=0$) consists of two parabolas that are shifted in momentum space according to the spin ($\ket{\rightarrow}$ or $\ket{\leftarrow}$ in the basis of $\sigma_x$ eigenstates). The Zeeman term $E_{\rm Z}\sigma_z$ does not commute with $\sigma_x$ and therefore opens a gap in the intersection points of the parabolas. Therefore, the position of the chemical potential $\mu$ determines whether there are two or four Fermi points. In the case of two Fermi points, introducing $\Delta\neq0$ opens an inverted superconducting gap, making the system a topological superconductor. In terms of the model parameters, the topological region appears at $E_{\rm Z}>\Delta$, i.e., the Zeeman energy must exceed the pair potential~\cite{lutchyn_majorana_2010,oreg_helical_2010}.

Let us now add current into the mix. The only change in the Hamiltonian compared to Eq.~\eqref{eq:majorana-nanowire} comes from the spatial variation of the SC phase $\phi(x)$, which changes the pairing terms to~\cite{romito_manipulating_2012-1} 
\begin{eqnarray}\label{eq:nanowire-current}
    {\cal H}_{\rm SC} = \Delta\left[\tau_x\cos\phi(x) +\tau_y\sin\phi(x) \right].
\end{eqnarray}
To make progress, it is very useful to introduce the gauge transformation 
\begin{eqnarray}\label{eq:gauge-current}
    {\cal U} = \exp\left[i\frac{\phi(x)}{2}\tau_z\right].
\end{eqnarray}
This transformation eliminates the phase variation from the pairing terms, at the cost of shifting the momentum in the normal-state terms, $p\rightarrow p-\tau_z\nabla\phi/2$. This seemingly innocent gauge transformation actually plays a very important role in understanding the mechanism by which phase variation changes the picture, as it establishes the similarity between a phase gradient and a vector potential. Notice that, as in the case of a real vector potential, the momentum shift is opposite for electrons and holes.

\begin{figure}
    \centering
    \includegraphics[width=\linewidth]{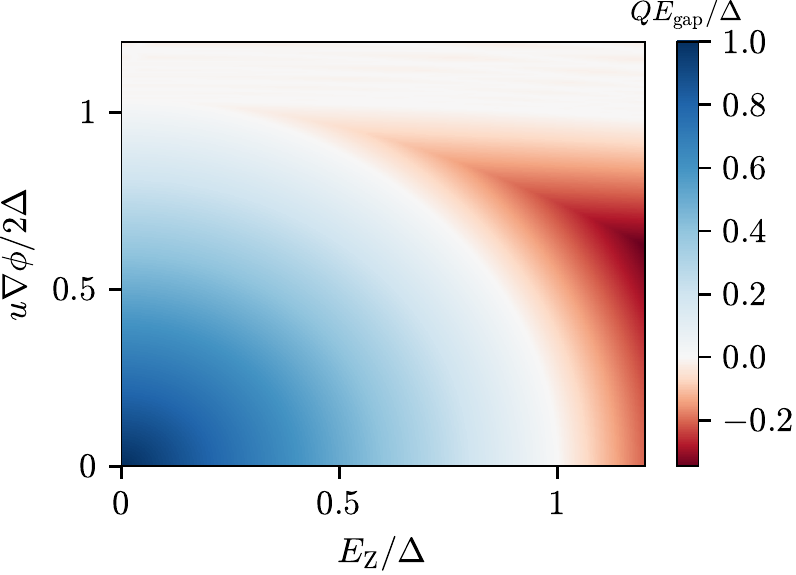}
    \caption{Topological phase diagram of the current-carrying wire, Eq.~\eqref{eq:nanowire-current} (see Ref.~\cite{romito_manipulating_2012-1}). As a function of the Zeeman field $E_{\rm Z}$ and the phase gradient $\nabla\phi$, the diagram shows the topological invariant ${\cal Q}$, which is $+1$ ($-1$) for the trivial (topological) phases, multiplied by the energy gap. Small phase gradient reduces the critical Zeeman field required for a topological transition, but large phase gradient makes the system gapless.}
    \label{fig:current_wire_PD}
\end{figure}

For the case of a constant current we get a constant phase gradient, and therefore the momentum shift is position independent. Then, the Hamiltonian may still be analyzed in momentum space (translation invariance is restored), and the topological phase boundaries are found by the closing of the gap at $p=0$. The topological phase diagram is shown in Fig.~\ref{fig:current_wire_PD}. Remarkably, the phase gradient gives rise to a topological phase at $E_{\rm Z}<\Delta$, which would not be possible in its absence. Notice, however, that as the current increases the superconducting gap decreases, until it eventually vanishes (see also Ref.~\cite{gennes_superconductivity_1999} for an elementary treatment of the gap's dependence on the phase gradient). This makes it impossible to tune into a topological phase at $E_{\rm Z}=0$.

\subsection{Planar Josephson junctions}\label{subsec:planarJJ}
Rather than applying a current to continuously vary the SC phase, it is possible to phase bias several SCs such that a discrete collection of phases is formed. The properties of the Josephson junctions formed between two such SCs strongly depend on the phase bias. As we show below, this dependence can greatly assist in engineering topological SC.

The first idea for utilizing Josephson junctions as Majorana platforms is due to Fu and Kane~\cite{fu_superconducting_2008}, who pointed it out in the same work reviewed in Sec.~\ref{sec:TISC}. There, two SCs are deposited on top of a topological insulator, forming an STIS junction. When the separation between the SCs is much smaller than $v/\Delta$ (where $v$ is the Fermi velocity and $\Delta$ is the SC gap), the junction can be considered as quasi 1D, and the dispersion is approximated as 
\begin{equation}\label{eq:STIS-dispersion}
    E(k)=\pm\sqrt{v^2 k^2 + \Delta^2 \cos^2\left(\frac{\phi}{2}\right)},
\end{equation}
where $\phi$ is the phase difference between the SCs. Crucially, at $\phi=\pi$ the system becomes gapless, which motivates the study of a low-energy effective theory near $\phi=\pi$, $k=0$. The effective Hamiltonian reads 
\begin{equation}\label{eq:STIS-Heff}
    \tilde{\cal H} = -i\tilde{v}\nu_x\partial_x + \tilde{\Delta}\nu_y,
\end{equation}
where the effective Fermi velocity $\tilde{v}$ depends on the chemical potential $\mu$ and the junction's width $W$, and the renormalized pair potential $\tilde{\Delta}=\Delta\cos(\phi/2)$. The Pauli matrices $\nu$ act in the two-dimensional low-energy subspace. The effective Hamiltonian has the same form as the Su-Schrieffer-Heeger (SSH) model~\cite{su_solitons_1979} and the Kitaev chain~\cite{kitaev_unpaired_2001}, and therefore at $\tilde{v}=\tilde{\Delta}$ it undergoes a topological phase transition, accompanied by the appearance of localized Majorana end states. However, it should be notes that the surface states at the sides of the 3DTI may mix with the MZMs, which is clearly detrimental in practice. An attempt to overcome this problem was made by Potter and Fu~\cite{potter_anomalous_2013}, utilizing Josephson vortices.

The STIS junction scheme is remarkably simple and elegant, but it is very challenging to realize experimentally, due to the difficulty in establishing a stable superconducting proximity effect in a 3DTI surface. To overcome this difficulty, two theoretical works~\cite{hell_two-dimensional_2017,pientka_topological_2017} proposed to exchange the 3DTI with a standard spin-orbit-coupled 2D electron gas (2DEG). This trading is possible since the Aharonov-Casher phase~\cite{aharonov_topological_1984}, arising from the Rashba spin-orbit coupling is actually the same as the $\pi$ Berry phase associated with the Dirac cone in the TI.

\begin{figure}
    \centering
    \includegraphics[width=\linewidth]{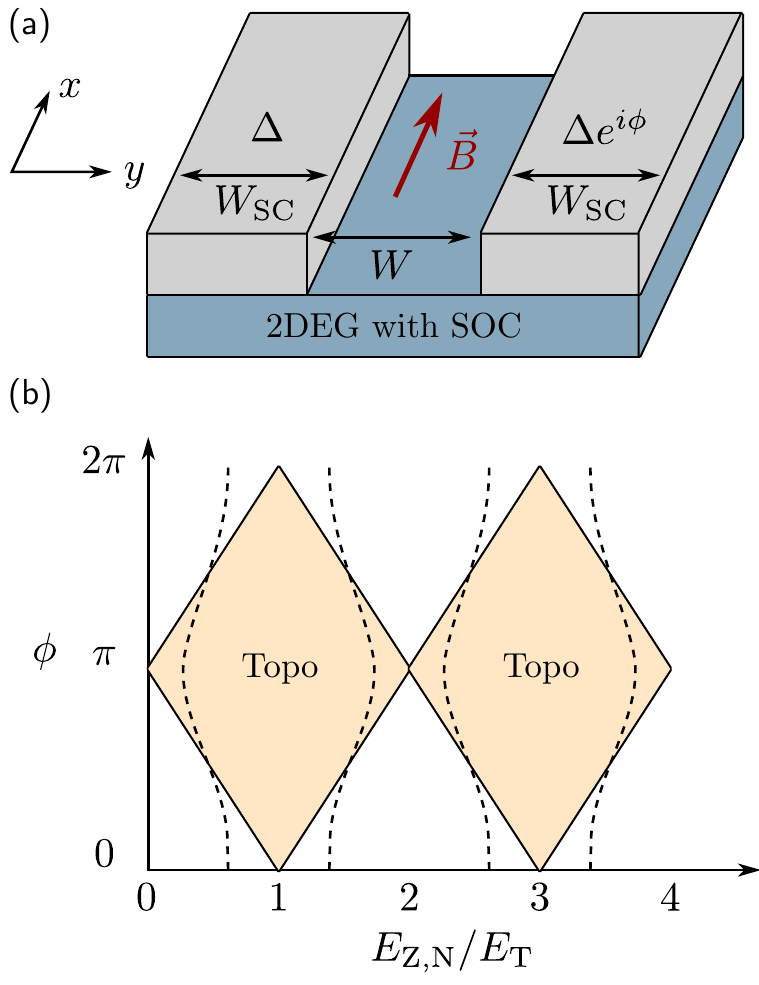}
    \caption{Topological superconductivity in a planar Josephson junction formed in a semiconductor-superconductor heterostructure (see Refs.~\cite{hell_two-dimensional_2017,pientka_topological_2017}. (a)~Model schematic: two superconductors (grey) of width $W_{\rm SC}$ are deposited on top of a 2D electron gas (blue). The width of the normal region separating the superconductors in $W$, and the phase difference between them is $\phi$. A magnetic field $B$ is applied in the $x$ direction in the junction's plane. (b)~Topological phase diagram as a function of the Zeeman energy in the normal region $E_{\rm Z,N}$ and the phase difference $\phi$. At $\phi=0$ the system becomes gapless when $E_{\rm Z,N}=E_{\rm T}$, with $E_{\rm T}=\pi v_{\rm F}/W$ the Thouless energy. At $\phi=\pi$ the system is gapless already at $E_{\rm Z,N}=0$. The topological region occupies the orange diamonds (marked by solid lines) in the case of perfect Andreev reflection at the interface, or the dashed regions in the presence of normal reflection.}
    \label{fig:planarJJ}
\end{figure}

Consider the setup depicted in Fig.~\figref{fig:planarJJ}{a}. A Josephson junction of phase difference $\phi$ is formed on top of a spin-orbit-coupled 2DEG, and an in-plane magnetic field is applied. The Hamiltonian describing the system is 
\begin{eqnarray}\label{eq:HPK}
    {\cal H}_{\rm JJ} = &\left[ -\frac{\partial_x^2+\partial_y^2}{2m} - \mu + i\alpha\left(\partial_x \sigma_y - \partial_y \sigma_x \right) \right]\tau_z\\
    &+ E_{\rm Z}\sigma_x + \Delta(y)\left[\tau_x\cos\phi(y) + \tau_y\sin\phi(y) \right]. \nonumber
\end{eqnarray}
Here $\alpha$ is the Rashba SOC strength, and $E_{\rm Z}=g\mu_{\rm B}B_{\parallel}/2$ is the Zeeman energy, where $g$ is the g-factor, $\mu_{\rm B}$ is the Bohr magneton, and $B_{\parallel}$ is the in-plane magnetic field, which we take to be in the $x$ direction. The SC pairing amplitude is $\Delta(y)=\Delta \Theta(|y|-W/2)$ and the phase is 
\begin{equation}
\phi(y)=\cases{
0&for $y < 0$\\
\phi&for $y>0$.\\}
\end{equation}

Some insight into the topological behavior of the system may be gained by analyzing the Hamiltonian~\eqref{eq:HPK} using the scattering formalism at $k_x=0$~\cite{pientka_topological_2017}. This is done by writing the expression for the Andreev reflection at the NS interface, 
\begin{equation}\label{eq:HPK_r_A}
    r_{\rm A}(E) = \exp\left( i\cos^{-1}\frac{E-E_{\rm Z,S}}{\Delta} + i\frac{\phi}{2} \right),
\end{equation}
and the expression for the electrons and holes transmission in the normal region,
\begin{equation}\label{eq:HPK_t_eh}
    t_{e(h)}(E) = \exp\left( \pm ik_{\rm F}{W} + i\frac{E-E_{\rm Z,N}}{v_{\rm F}}W \right).
\end{equation}
Here $E_{\rm Z,S},E_{\rm Z,N}$ are the Zeeman energies in the superconducting and normal regions, respectively (these differ in general due to different g-factors in the different regions), and $W$ is the junction's width. In the absence of normal reflection at the NS interface, the condition for a bound state at energy $E$ is $t_e(E) t_h(E) r_{\rm A}^2(E)=1$. For $E_{\rm Z,S}=0$, a zero-energy state, signaling a topological phase transition (since we are looking at $k_x=0$), appears for
\begin{equation}\label{eq:HPK_boundaries}
    \frac{E_{\rm Z,N}}{E_{\rm T}} \pm \frac{\phi}{\pi} = 1 + 2n,
\end{equation}
where $E_{\rm T}=\pi v_{\rm F}/2W$ is the Thouless energy. This diamond-like phase diagram is shown in Fig.~\figref{fig:planarJJ}{b}. Including also normal reflection and $E_{\rm Z,S}\neq0$ smears the sharp edges of the diamonds. Notice that even though the SOC does not appear in Eq.~\eqref{eq:HPK_boundaries}, its presence is essential to the existence of the topological phase: without SOC, there would be no gap in the topological phase.

One of the important advantages of this scheme is its weak sensitivity to the chemical potential. Moreover, misalignment of the magnetic field in the plane has no dramatic effects, making it in principle more robust than nanowires. Importantly, the energy scale to which the Zeeman energy should be compared is the Thouless energy $E_{\rm T}\propto1/W$. This sets the scale for a topological transition near $\phi=0$, but it is also relevant at larger value of $\phi$. The fact that $E_{\rm T}$ is inversely proportional to $W$ motivates the use of wide junctions, since those would require small magnetic fields to become topological. However, narrow junctions are generally preferred as they allow quasi-ballistic motion. These two competing effects are the experimental challenges in realizing the planar Josephson junction scheme. 

Two experiments attempting to realize this planar proposal are worth mentioning at this point. Ren \etal~\cite{ren_topological_2019} used a HgTe quantum well as the 2DEG (which has the advantage of very strong spin-orbit coupling), and thermally evaporated thin-film Al on top of it. Fornieri \etal~\cite{fornieri_evidence_2019} used InAs as the 2DEG (which has the advantage of high mobility), with a thin Al layer epitaxially grown on top. The two experiments reported zero-bias conductance peaks consistent with topological superconductivity. Curiously, the experiments were performed with different material platforms and at quite different length regimes (narrow vs. wide junction, narrow vs. wide SC leads compared to the SC coherence length), but they both showed encouraging signatures of MZMs.

As a closing remark, we touch upon a delicate issue which is of primary importance in experiments. The model of Refs.~\cite{hell_two-dimensional_2017,pientka_topological_2017} indeed predicts a topological phase, but its topological gap is only about a third of the induced SC gap, at best. The culprit turns out to be long semiclassical trajectories that do not scatter off SCs, and thus lead to a small gap at finite $k_x$. One way to get rid of these trajectories is impurities, which break the trajectories and make them scatter onto the SCs too. Indeed, it was found that the right amount of disorder may increase the gap and shorten the Majorana localization length~\cite{haim_benefits_2019}. A different approach is changing the geometry of the junction such that these trajectories naturally scatter, for example by using a zig-zag shape~\cite{laeven_enhanced_2020}. This was found to confine the MZM wavefunction more compared to the straight junction case.

\section{Topological superconductivity induced by phase bias only}\label{sec:phaseonly}
Having reviewed proposals on reducing the critical magnetic field, we now move to ideas on eliminating the Zeeman field altogether. The systems surveyed in this section aim to use just the SC phase to obtain topological superconductivity, but they do so in rather different ways. We note that early ideas along these lines appeared in the work of Kotetes~\cite{kotetes_topological_2015}, who suggested several configurations where flux or supercurrent can eliminate the need for a Zeeman field.

\subsection{Full-shell nanowires}
The original models of semiconducting nanowires~\cite{lutchyn_majorana_2010,oreg_helical_2010} included the effect of the magnetic field only via the Zeeman splitting. In other words, the nanowires were treated as truly one-dimensional objects, so orbital effects were neglected. In reality, however, nanowires have a finite cross section so flux can be threaded through them. This important effect is at the core of the full-shell nanowires scheme proposed by Vaitiekėnas \etal~\cite{vaitiekenas_flux-induced_2020}.

\begin{figure}
    \centering
    \includegraphics[width=\linewidth]{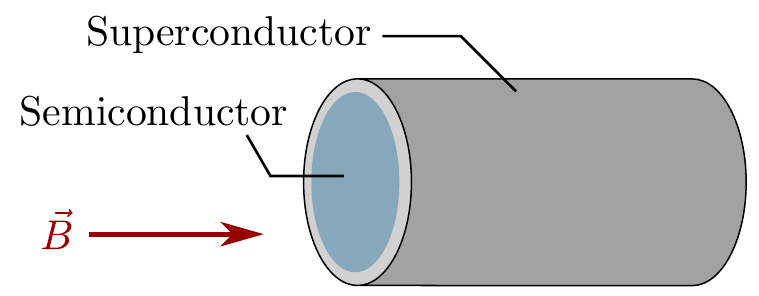}
        \caption{Full-shell nanowire scheme for topological superconductivity~\cite{vaitiekenas_flux-induced_2020}. A cylindrical semiconducting wire (blue) is fully coated by a superconductor (grey), and a magnetic field $\vec{B}$ is applied along the wire's axis. The orbital effect of the magnetic field, and in particular the phase bias it induces in the superconductor, drives the system into a topological phase.}
    \label{fig:fullshell}
\end{figure}

Consider a cylindrical nanowire (e.g. InAs) fully coated by a superconductor (e.g. Al), see Fig.~\ref{fig:fullshell}. The spin-orbit coupling in the nanowire arises from local breaking of inversion symmetry, and crucially the effective electric field points in the radial direction $\hat{r}$. The continuum normal-state Hamiltonian is then
\begin{equation}
    H_0 = \frac{\left(\vec{p}-e\vec{A}\right)^2}{2m} - \mu + \alpha\hat{r}\cdot\left[ \vec{\sigma}\times \left(\vec{p}-e\vec{A}\right) \right].
\end{equation}
Here $\vec{A}$ is the vector potential, $\alpha$ is the Rashba spin-orbit coupling strength, and $\vec{\sigma}$ is the vector of Pauli matrices acting in spin space. We assume a magnetic field $\vec{B}=B\hat{z}$ is applied along the nanowire's axis, so a reasonable gauge choice is $\vec{A}=\frac{1}{2}\vec{B}\times\vec{r}$.

On top of this normal-state Hamiltonian, we add the superconducting pairing, which is modulated in space due to the presence of the magnetic field: $\Delta\left(\vec{r}\right)=\Delta e^{-in\phi}$, where $\phi$ is the angular coordinate. The winding $n$ is strictly an integer, due to the wavefunction being single valued. An important observation of Vaitiekėnas \etal~\cite{vaitiekenas_flux-induced_2020} is that the BdG Hamiltonian commutes with the generalized angular momentum operator
\begin{equation}\label{eq:fullshell_J_z}
    J_z=-i\partial_{\phi} + \frac{1}{2}\sigma_z + \frac{n}{2}\tau_z.
\end{equation}
The first two terms are not surprising---they are nothing but the orbital and spin angular momenta. The last term, however, is special: the phase winding alters the angular momentum and adds a contribution that is opposite for particles and holes. This is yet another manifestation of the notion (that has hopefully become more transparent at this point) that SC phase gradient can be thought of as a special type of orbital field.

Inspecting Eq.~\eqref{eq:fullshell_J_z}, we see that the eigenvalue~$m_J$ of the orbital angular momentum $-i\partial_{\phi}$ takes on integer values of odd~$n$ and half-integer values for even~$n$. Since the particle-hole operator (which by necessity anti-commutes with the BdG Hamiltonian) relates a state with energy $E,m_J$ to a state with $-E,-m_J$, it is evident that at the $m_J=0$ sector there is hope for non-degenerate Majorana zero modes. That is because the particle-hole symmetry at $E=0,m_J=0$ relates a state with itself and therefore does not add any degeneracy. It is therefore worthwhile to examine $m_J=0$, and that can only be done at odd $n$ for this system. The conclusion is that an odd winding---one vortex at least---is needed for a topological transition.

Following this line of reasoning, the authors of Ref.~\cite{vaitiekenas_flux-induced_2020} then apply a unitary transformation to the BdG Hamiltonian to bring it to a form similar to the original nanowire Hamiltonian~\cite{lutchyn_majorana_2010,oreg_helical_2010}, see Eq.~\eqref{eq:majorana-nanowire}. We do not repeat the full expressions here, but the key point is that at $m_J=0$ the model maps exactly to the Majorana nanowire Hamiltonian, with a chemical potential shifted by the flux and the spin-orbit coupling, and an \emph{emergent} Zeeman field,
\begin{equation}
    E_{\rm Z} = \left(n-\frac{\Phi}{\Phi_0}\right) \left( \frac{1}{4mR^2} + \frac{\alpha}{2R} \right),
\end{equation}
where $R$ is the cylinder's radius (assuming a thin superconductor) and $\Phi=\pi R^2 B$ is the flux threaded through the cylinder. It is thus straightforward to derive the criterion for a topological phase in this description, and indeed the authors found large topological regions in parameter space. We stress again that no Zeeman field was assumed to be present in the original Hamiltonian, and it is indeed the SC phase winding that gives rise to the effective Zeeman field.

The relevant magnetic field scale for a topological transition is determined by $\Phi\approx\Phi_0$, i.e., by the geometry of the nanowire. Notice again that this is a purely orbital effect (no $g$ factors involved). In practice, for the Al-coated InAs nanowire studied in Ref.~\cite{vaitiekenas_flux-induced_2020}, this field scale is around 100~mT (significantly lower than the field scale for a Zeeman-driven transition in proximitized InAs nanowires~\cite{mourik_signatures_2012,das_zero-bias_2012}). The zero-bias peaks, which may imply the existence of Majorana zero modes, are observed experimentally as additional features on top of the Little-Parks oscillations.

We note that a similar phenomenon of orbitally driven topological superconductivity has been pointed out in carbon nanotubes~\cite{lesser_topological_2020}. These are sheets of graphene rolled up to form quasi-1D tubes, and due to their finite diameter, they can also capture magnetic flux. Since they are made entirely of carbon atoms, carbon nanotubes have an extremely small $g$ factor, and therefore the orbital effect can in fact be the dominant one. However, the band structure and spin-orbit coupling in carbon nanotubes are markedly different compared to semiconducting nanowires such as InAs, making the relevant physics quite different. We note that Zeeman-based schemes in CNTs have also been proposed~\cite{sau_topological_2013,marganska_majorana_2018}, but they require very large magnetic fields.

\subsection{Supercurrent in a planar Josephson junction}
The realization that phase bias can be an alternative to applying a Zeeman field was also adopted in the context of planar Josephson junctions (see Sec.~\ref{subsec:planarJJ} and Refs.~\cite{hell_two-dimensional_2017,pientka_topological_2017}) by Melo~\etal~\cite{melo_supercurrent-induced_2019}. Recall that the original proposal for topological superconductivity~\cite{hell_two-dimensional_2017,pientka_topological_2017} already has phase bias, but only in one direction ($y$ in the notation of Fig.~\ref{fig:planarJJ}).
Melo~\etal~\cite{melo_supercurrent-induced_2019} proposed to extend this to a continuous phase gradient along the junction ($x$ in the notation of Fig.~\ref{fig:planarJJ}), thereby potentially eliminating the need for a Zeeman field in this direction. 
The need for a phase gradient in both dimensions becomes clear when considering $k_x=0$. In the absence of phase gradient, a $y$-dependent gauge transformation removes the spin-orbit coupling. In the absence of a Zeeman field in the $x$ direction, this makes all states doubly degenerate, thus prohibiting a transition into the topologically non-trivial phase.

The first step taken by Melo~\etal is applying supercurrents to the two SCs, generally with different gradients:
\begin{equation}
\phi(y)=\cases{
\exp\left(2\pi i x/\lambda_{\rm L}\right)&for $y < 0$\\
\exp\left(-2\pi i x/\lambda_{\rm R}\right)&for $y>0$.\\}
\end{equation}
Notice that the effective unit cell of the junction is now enlarged: it is the least common multiple of $\lambda_{\rm L}$ and~$\lambda_{R}$. While these supercurrents lift the degeneracy at $k_x=0$, the spectrum turns out to be gapless due to a gap closing at finite $k_x$. This gap closing is a result of a symmetry dubbed ``charge-momentum parity", which reflects the fact that Andreev reflections in this system are always accompanied by a momentum shift of $\pm2\pi/\lambda_{\rm L}$ or $\pm2\pi/\lambda_{\rm R}$. This symmetry, ${\cal O}=(-1)^n \tau_z$ (where $n$ is the number of unit cells in reciprocal space), which commutes with the Hamiltonian and anti-commutes with the particle-hole operator, prohibits a gapped topological phase.

Several mechanisms are proposed to break the charge-momentum parity:
\begin{enumerate}
    \item Adding a periodic potential, which scatters states of different wavevectors inside the electron and hole sectors separately. 
    \item Deforming the junction to a non-straight shape such as zig-zag. Here, the geometry makes the Andreev reflection change $k_x$ in a way that breaks~$\cal O$.
    \item Adding a third SC without supercurrent in the middle of the junction (creating a SNSNS junction). The additional SC couples states at $k_x$ and $-k_x$ without any momentum transfer.
\end{enumerate}
Interestingly, in cases (i)--(ii), modulations commensurate with $\lambda_{\rm L}$ and $\lambda_{\rm R}$ are favorable for breaking ${\cal O}$ and opening a topological gap, whereas incommensurate modulations facilitate a gapless state. This is because in the commensurate case---and always in case (iii) above if $\lambda_{\rm R}=\lambda_{\rm L}$---inversion symmetry is maintained. When inversion symmetry is broken, the spectrum gets tilted and can easily become gapless, and thus retaining inversion symmetry is beneficial in this setup.

An important lesson to learn from Ref.~\cite{melo_supercurrent-induced_2019} is that SC phase winding can be a sufficient time-reversal-symmetry breaking mechanism, but the directions matter a lot. If time-reversal is broken in such a way that the effective Zeeman field is parallel to the spin-orbit coupling, the spectrum simply becomes gapless. In fact, even if the SC phase changes in all directions, one still needs clever tricks to avoid gap closings.

\subsection{Discrete vortices in superconductor-semiconductor heterostructures}
A different approach to phase-only Majorana realization, introduced in Refs.~\cite{lesser_three-phase_2021,lesser_phase-induced_2021}, utilizes the concept of a discrete vortex (see Sec.~\ref{sec:TISC}) in conjunction with spin-orbit coupling. Here we review the basic idea, and two practical routes for implementation.

\begin{figure}
    \centering
    \includegraphics[width=\linewidth]{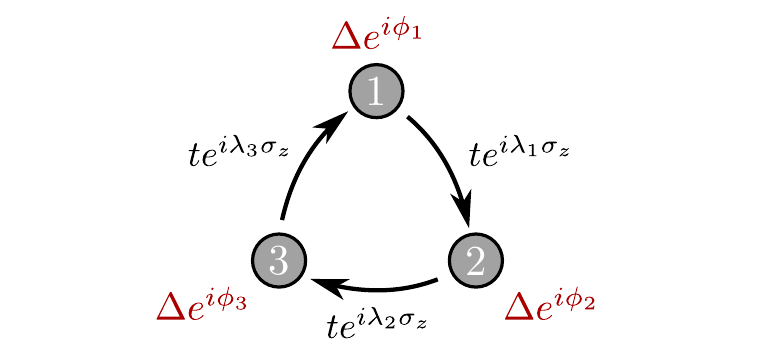}
    \caption{Ring with three phases and spin-orbit coupling.}
    \label{fig:ring_phases}
\end{figure}

The analysis begins with a spin-orbit-coupled quantum ring, where each site is coupled to a superconductor which has a different phase, see Fig.~\ref{fig:ring_phases}.
Form Ref.~\cite{van_heck_single_2014} we know that three different superconducting phases are necessary to have a zero-energy Andreev state.
Since we will be interested in forming zero-energy states, the minimal number of sites in the ring is three.

The Hamiltonian describing the ring is~\cite{lesser_three-phase_2021,lesser_phase-induced_2021}
\begin{eqnarray}\label{eq:ring_phases}
    H = &-\sum_{n=1}^{3} \mu c^{\dagger}_n c_n \\
    &+ \sum_{n=1}^{3}\left( -c^{\dagger}_n te^{i\lambda_n \sigma_z} c_{n+1} + \Delta e^{i\phi_n} c^{\dagger}_{n\uparrow} c^{\dagger}_{n\downarrow} + {\rm H.c.} \right). \nonumber
\end{eqnarray}

Here $n$ enumerates the three sites of the ring, $c_n=\left( c_{n\uparrow}, c_{n\downarrow} \right)^{\rm T}$ and the phases $\lambda_n$ characterize the spin-orbit coupling strength at each bond. The reader might be concerned about the SOC being always related with the $\sigma_z$ operator and not the other spin operators; a seemingly more appropriate operator might be $\sigma_{\theta}=-\sigma_{x}\sin\theta+\sigma_{y}\cos\theta$, where $\theta$ is the angular polar coordinate. The two representations are actually related by the local transformation $\mathcal{U}=\exp\left[\frac{i}{2}\left(\theta+\frac{\pi}{2}\right)\sigma_{z}\right]$ (see also Ref.~\cite{vaitiekenas_flux-induced_2020}): $\mathcal{U}\sigma_{\theta}\mathcal{U}^{\dagger}=\sigma_{x}$ and $\sigma_x$ may easily be rotated into $\sigma_z$. The SC part of the Hamiltonian, being a spin-singlet term, is invariant under such a rotation, thereby justifying the use of the simpler SOC term appearing in Eq.~\eqref{eq:ring_phases}, by essentially redefining $c_n\to {\cal U}^{\dagger} c_n{\cal U}$.

Let us first consider the simple case of uniform SOC, $\lambda_1=\lambda_2=\lambda_3\equiv\lambda$, and equal SC phase differences, $\phi_n=\frac{2\pi}{3}mn$ where $m=\pm1$. In this case the model has translation invariance and it is conveniently analyzed in Fourier space, using the momentum around the ring $k$ as a conserved quantum number. The Bloch-BdG Hamiltonian is 
\begin{eqnarray}
    \mathcal{H}\left(k\right) = \left( \begin{array}{cccc}
        \epsilon_{k\uparrow} & 0 & \Delta & 0\\
        0 & \epsilon_{k\downarrow} & 0 & \Delta\\
        \Delta & 0 & -\epsilon_{-k\downarrow} & 0\\
        0 & \Delta & 0 & -\epsilon_{-k\uparrow}
    \end{array} \right),
\end{eqnarray}
where $\epsilon_{k\sigma}=-\mu+2t\cos\left(k+\frac{\pi m}{3}+\lambda\sigma\right)$ and as usual the Nambu spinor takes the form $\vec{\Psi}_{k}=\left(c_{k\uparrow},c_{k\downarrow},c_{-k\downarrow}^{\dagger},-c_{-k\uparrow}^{\dagger}\right)^{\rm T}$. To find zero-energy states, we demand that for some value of $k$, 
\begin{equation}
    \det\mathcal{H}\left(k\right)=\left(\Delta^{2}+\epsilon_{k\uparrow}\epsilon_{-k\downarrow}\right)\left(\Delta^{2}+\epsilon_{k\downarrow}\epsilon_{-k\uparrow}\right)=0.
\end{equation}
There are many ways to fulfil this equation. Let us focus on a case that simplifies the calculation: $\epsilon_{k\sigma}=-\epsilon_{-k,-\sigma}=\Delta$ for some specific choice of $k,\sigma$. This leads to two equations (see Ref.~\cite{lesser_phase-induced_2021} for additional details), that can be further simplified by setting $\mu=0$. This leads to the requirement $\Delta=\pm\sqrt{3}t$, and to a requirement on the SOC parameter (for $m=1$):
\begin{eqnarray}
    k + \lambda\sigma + \frac{\pi m}{3} = \pm\frac{\pi}{6} + 2\pi N,
\end{eqnarray}
where $N$ is an integer. For example, the choice $k=0$, $\sigma=\uparrow$ leads to $\lambda=-\frac{\pi}{2}+2\pi N$. We therefore conclude that it is possible to induce zero-energy states, which are delocalized Majorana states, using just three phases. Notice the interplay between $\lambda$, the Aharonov-Casher phase~\cite{aharonov_topological_1984}, and $m$, the SC phase winding.

The analysis may be generalized to a ring without translation invariance~\cite{lesser_three-phase_2021}. In this case, the determinant of the full BdG Hamiltonian~\eqref{eq:ring_phases} may be written explicitly:
\begin{eqnarray}\label{eq:det_H_ring}
\det H&=6\mu^{2}t^{2}\left(\Delta^{2}+\mu^{2}\right) -\left(\Delta^{2}+\mu^{2}\right)^{3} \\
&-3t^{4}\left(\Delta^{2}+3\mu^{2}\right) -2f\Delta^{2}t^{2}\left(\Delta^{2}+\mu^{2}+t^{2}\right) \nonumber \\
&-4\mu t^{3}\Lambda\left(f\Delta^{2}-\mu^{2}+3t^{2}\right) -4t^{6}\Lambda^2, \nonumber
\end{eqnarray}
where $f=\cos\left(\phi_{1}-\phi_{2}\right)+\cos\left(\phi_{2}-\phi_{3}\right)+\cos\left(\phi_{3}-\phi_{1}\right)$ and $\Lambda=\cos\left(\lambda_1+\lambda_2+\lambda_3\right)$. To find zero-energy states, we look for solutions of the equation ${\rm det}H=0$; notice that these only depend on the total spin-orbit phase (characterized by $\Lambda$), and the SC phases enter through a single gauge-invariant parameter $f$. 

Treating Eq.~\eqref{eq:det_H_ring} as a quadratic equation for $\Lambda$, it is readily understood that regardless the values of $t,\,\mu,$ and $\Delta$, real solutions of $\Lambda$ are possible only when $f\le -1$. It can readily be shown~\cite{lesser_three-phase_2021} that $f$ is directly related to phase winding: a discrete vortex is formed if and only if $f\leq-1$. Therefore, we find that phase winding is a \emph{necessary} condition for a zero-energy state.
Continuing the analysis of the quadratic equation, one can then find a set of \emph{optimal} values of $\mu/t,\Delta/t,\Lambda$, such that $f\leq-1$ becomes a \emph{sufficient} condition: for these values, any $-3/2\leq f\leq-1$ will induce a zero-energy state. The manifold in this three-dimensional parameter space turns out to be the circle $\Delta=\sqrt{t^2-\mu^2}$, $\Lambda=\mu/t$, and on this circle we say that $f_{\rm crit}=-1$.
This criterion may be translated to continuum parameters, leading to the rule of thumb $L\Delta\approx\alpha$, where $L$ is the typical length of the system and $\alpha$ is the spin-orbit coupling constant.
Moving away from this optimal circle reduces $f_{\rm crit}$ (notice that smaller $f_{\rm crit}$ is a more stringent condition on the phases; the minimal value $f$ may take is $-3/2$ and it corresponds to a perfect vortex).

At this point, we would like to extend this ring to a 1D system, building on the observations above. One can think of two ways to do that, which can be regarded as different ways to concatenate rings. The first way we describe amounts to coupling the rings out of the plane, forming a cylinder-like structure, as described in Ref.~\cite{lesser_three-phase_2021}. The other way, which is explored in Ref.~\cite{lesser_phase-induced_2021} is repeating the rings in the plane.

\begin{figure}
    \centering
    \includegraphics[width=\linewidth]{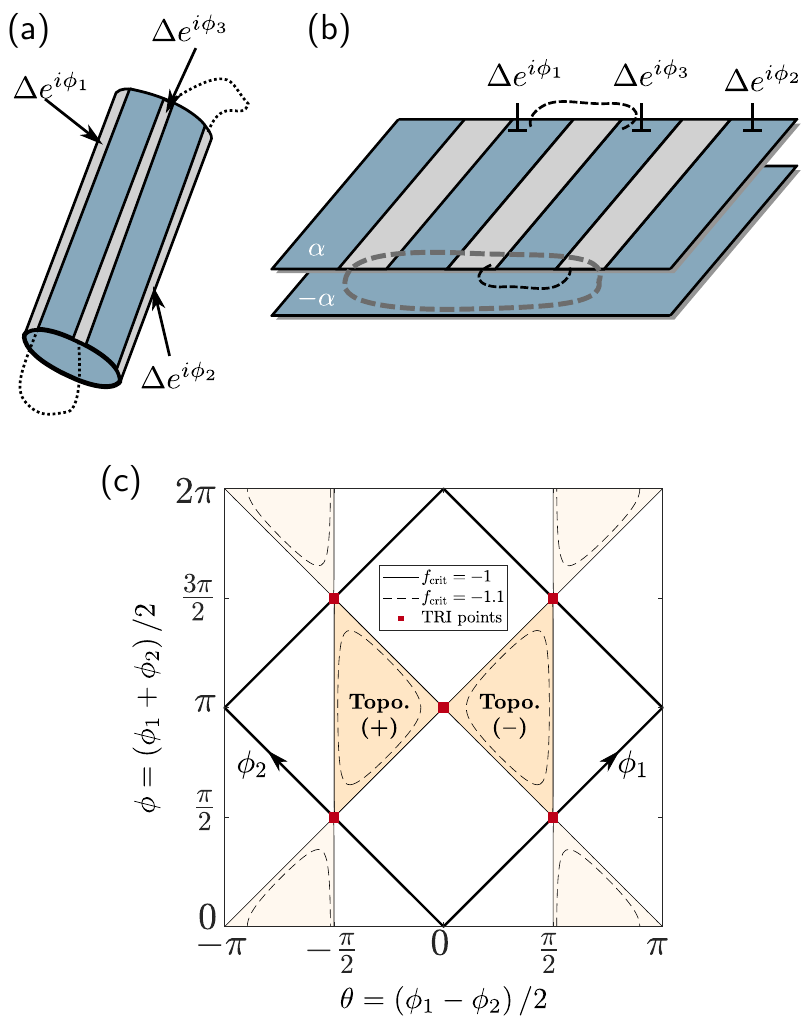}
    \caption{Non-planar proposal for phase-only MZMs (see Ref.~\cite{lesser_three-phase_2021}. (a)~Each site of the ring of Fig.~\ref{fig:ring_phases} is promoted to a wire. (b)~Deformation of the cylinder to a bilayer semiconductor 2DEG with three SCs deposited on top. (c)~Topological phase diagram of the cylinder model (a), as a function of the two phase differences. For optimal parameters, $f_{\rm crit}=-1$ and the topological phase occupies the full orange triangles. For sub-optimal parameters, $f_{\rm crit}<-1$ and the phase boundaries are curved.}
    \label{fig:threephase}
\end{figure}

Consider promoting each site of the ring of Fig.~\ref{fig:ring_phases} to be a wire, as depicted in Fig.~\figref{fig:threephase}{a}. The ring's Hamiltonian Eq.~\eqref{eq:ring_phases} is then understood as the $k_{\parallel}=0$ Hamiltonian of the coupled wires system. Therefore, the zero-energy states of this Hamiltonian correspond to topological phase transitions. We can use the form of the determinant, Eq.~\eqref{eq:det_H_ring}, to understand the topological phase diagram. From our previous arguments we know that phase winding, or $f\leq-1$, is a necessary condition for a topological phase, and we also know when this becomes a sufficient condition. On the optimal manifold we found, any $-3/2\leq f<-1$ will drive the system into the topological phase, and the phase diagram will assume the form of triangles as in the Fu-Kane model (see Ref.~\cite{fu_superconducting_2008} and Fig.~\ref{fig:fu_kane}). Away from this optimal manifold, the phase boundaries get distorted: they are no longer straight lines, but rather contours of constant $f$ (see Fig.~\figref{fig:threephase}{c}).

An important consideration we need to account for is the topological gap. The system can only be topological if it has a finite gap, and the larger the gap the better: it provides the protection of the MZMs and determines their localization length. One may expect the gap to be largest when the phases form a perfect vortex ($f=-3/2$). However, it turns out that this is not the case: for such a configuration, the system has $C_3$ rotation symmetry, which manifests itself as a gap closing at \emph{finite} $k_{\parallel}$. Therefore, besides optimizing the topological region in parameter space, one should also be concerned with the non-trivial evolution of the gap. The optimal topological gap is numerically found to be about a third of the induced SC gap~\cite{lesser_three-phase_2021}.

This non-planar construction may be extended to a more realistic setup. Consider smoothly deforming the cylinder into a rectangular box, which we model as a bilayer, see Fig.~\figref{fig:threephase}{b}. The cylindrical geometry is manifested by a different SOC field in each layer. Another way to make this transition is to imagine turning off the periodic boundary conditions, so we are left with just three wires in the plane. The SOC cannot contribute in this scenario---we need to engineer a way for it to not cancel out. This is done by introducing an additional band in each wire, with a different SOC amplitude. Electrons can then travel in closed trajectories that capture a non-zero Aharonov-Casher phase; for example, the trajectory can go left-to-right in the ``top" layer, then hop down, return right-to-left, and finally hop back up. The advantage of this geometry is its experimental accessibility by means of simple semiconducting quantum wells, and indeed it has a very similar phase diagram to the cylinder model~\cite{lesser_three-phase_2021}.

The second approach to this phase-only setup is concatenating the rings of Fig.~\ref{fig:ring_phases} in the plane. A conceptually simple construction goes as follows: we take two rings and tune each of them to have zero-energy states. For simplicity, this can be done using the assumption of a perfect vortex (the details of the derivation in this case are given in Ref.~\cite{lesser_phase-induced_2021}, but notice that it is not essential to assume a perfect vortex). At this point, we have a total of four MZMs, two localized in each ring. Our goal is to turn on coupling between the rings such that two of these states are gapped out, and we are left with one MZM per ring. It turns out that to achieve this, the rings must be connected via more than one link (the physical reason for this is that eliminating two MZMs while leaving the other two intact is only possible via interference). The coupling may be chosen such that this elimination is perfect, leading to a so-called ``sweet spot", analogous to the Kitaev chain in the $t=\Delta$, $\mu=0$ limit. It is now possible to repeat this two-rings unit cell as many times a one desires, and regardless of how the unit cells are coupled, we will end up with one MZM localized at each end of the system. The conclusion is that phase bias can be used to drive the system into the most localized topological regime possible.

\begin{figure}
    \centering
    \includegraphics[width=\linewidth]{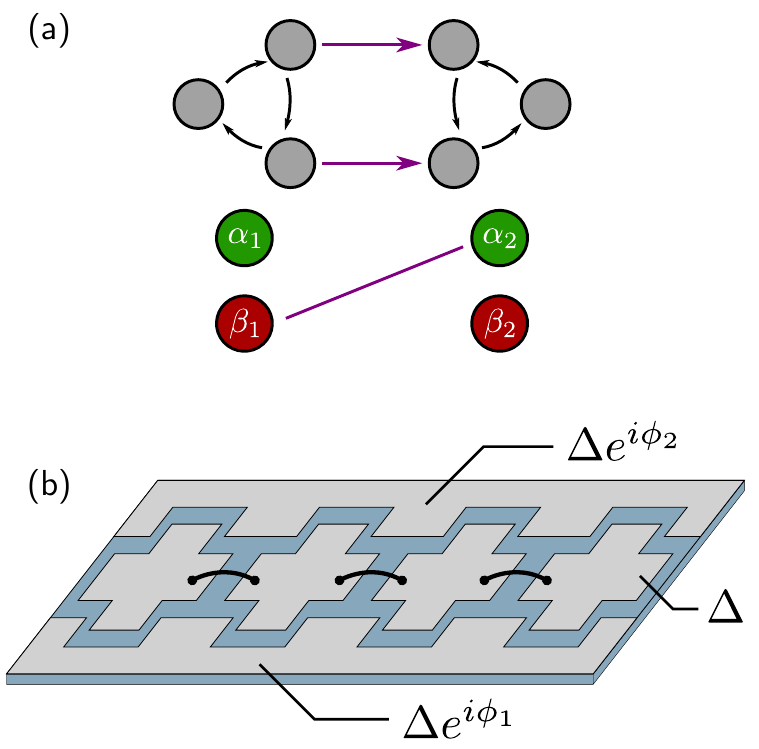}
    \caption{Planar proposal for phase-only MZMs (see Ref.~\cite{lesser_phase-induced_2021}. (a)~Two rings of the type shown in Fig.~\ref{fig:ring_phases} are tuned to have MZMs, and then coupled via two links (purple). If the coupling is chosen properly, only two ($\beta_1$, $\alpha_2$) of the four MZMs are gapped out, and each ring hosts an uncoupled MZM ($\alpha_1$ on the left, $\beta_2$ on the right). (b)~Realistic planar device geometry: three SCs (grey) are deposited on top of a spin-orbit-coupled 2DEG (blue). The SCs are phase biased, and the geometry enables electron trajectories encircling a non-vanishing spin-orbit phase and encountering SC phase winding.}
    \label{fig:phases-planar}
\end{figure}

As in the case of the non-planar model, this planar construction may also be adopted to a more experimentally feasible setup~\cite{lesser_phase-induced_2021}. A proposal for such a setup is shown in Fig.~\ref{fig:phases-planar}. The idea is forcing electrons to go in closed trajectories that pick up an Aharonov-Casher phase (due to Rashba SOC in a 2DEG) and are also sensitive to the phase bias. Due to the inherent asymmetry between the three phases in this proposal, it does not suffer from the problem of gap closing in the presence of a perfect vortex. Other than that, its topological phase diagram is very similar to that of the non-planar model.

\section{How to phase-bias a superconductor}\label{sec:howto}

In this section we briefly discuss the practical question of how one can achieve SC phase control. The basic theory has been well understood for a long time~\cite{tinkham_introduction_2004,gennes_superconductivity_1999}, and here we highlight some experimental aspects.

\begin{figure}
    \centering
    \includegraphics[width=\linewidth]{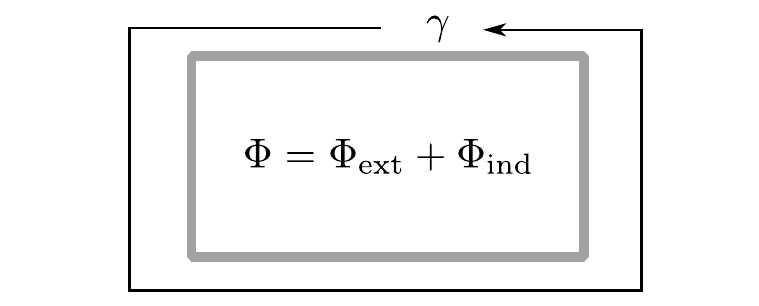}
    \caption{Phase control in a mesoscopic superconducting ring~\cite{majer_simple_2002}. The total flux $\Phi$ threading the ring is the sum of the externally applied flux $\Phi_{\rm ext}$ and the flux induced by the current flowing along the ring $\Phi_{\rm ind}=L_{\rm G}I$. The flux causes a phase difference $\gamma$ obeying the quantization condition $\gamma+2\pi\Phi=2\pi n$ where $n$ is an integer.}
    \label{fig:biasing}
\end{figure}

The basic setup~\cite{majer_simple_2002} consists of a mesoscopic superconducting ring, see Fig.~\ref{fig:biasing}. The ring is subjected to an external magnetic flux $\Phi_{\rm ext}$, as well as to a current $I$ which gives rise to an induced flux $\Phi_{\rm ind}=L_{\rm G}I$, where $L_{\rm G}$ is the ring's geometric inductance. The phase difference $\gamma$ across the ring's circumference follows the quantization condition
\begin{equation}\label{eq:gamma_ring}
    \gamma + 2\pi\phi = 2\pi n,
\end{equation}
where the integer $n$ is the fluxoid number and $\phi=(\Phi_{\rm ext}+\Phi_{\rm ind})/\Phi_0$. This integer is related to the presence of vortices in the SC: changing it requires the SC order parameter to go to zero in some region, through a phase-slip process. Since the phase difference also obeys $\gamma=2\pi L_{\rm K}I/\Phi_0$ where $L_{\rm K}$ is the loop's kinetic inductance, we find that $    \left( L_{\rm K}+L_{\rm G} \right)I = n\Phi_0-\Phi_{\rm ext}$. This immediately gives the $\Phi_{\rm ext}$-dependent energy of the ring for each $n$, 
\begin{equation}
    E = \frac{1}{2}\left( L_{\rm K}+L_{\rm G} \right)I^2 = \frac{\left( n\Phi_0-\Phi_{\rm ext} \right)^2}{2\left( L_{\rm K}+L_{\rm G} \right)}.
\end{equation}
This is a family of parabolas that intersect at half-integer value of $\Phi_{\rm ext}/\Phi_0$, which is when a vortex is able to enter the system.

It is possible to concentrate the phase drop across the ring to a short segment, by lowering the SC order parameter in this segment. Such ``weak link" configurations are often used in superconducting quantum interference devices~\cite{fagaly_superconducting_2006}.
We therefore find two ways to achieve phase bias between two superconductors: threading magnetic flux and applying an external current. In experiments involving a single phase difference~\cite{ren_topological_2019,fornieri_evidence_2019}, it is very common to use the former, by applying a weak out-of-plane magnetic field. Since $\Phi_{\rm ext}$ grows with the ring's area, it is easy to achieve a phase difference of order $\pi$ with extremely small magnetic fields. Setups involving more than one phase difference~\cite{lesser_three-phase_2021,lesser_phase-induced_2021} are more complicated to realize using just magnetic flux, since the loops' area cannot be controlled \emph{in situ}. Therefore, it might be preferable in these cases to use currents, thus allowing greater controllability during the experiment.

Another way to control the SC phase in a heterostructure is utilizing the \emph{Doppler shift} caused by an in-plane magnetic field~\cite{tkachov_magnetic_2004,fagas_geometrical_2005,rohlfing_doppler_2009}. Consider a 2DEG on top of which several SC electrodes are proximitized, each having a different height. Then, an in-plane magnetic field induces a spatially varying SC phase,
\begin{equation}
    \phi(\vec{r}) = -\frac{2\pi}{\Phi_0}\int_{\vec{r}_0}^{\vec{r}} \vec{A}(\vec{r'})\cdot d\vec{r'},
\end{equation}
where $\vec{A}(\vec{r})$ is the vector potential and $\vec{r}_0$ is some arbitrary reference point. Crucially, the relevant area to determine phase of each SC depends on its height, and this allows sensitive phase biasing. In the literature, this effect is usually seen via its influence on the supercurrent: the gradient of the SC phase, $\vec{k}_{\rm S}=\nabla\phi$, induces a diamagnetic current. This in turn means that the total momentum of a pair of quasiparticles undergoing Andreev reflection is $\vec{k}_1+\vec{k}_2=\vec{k}_{\rm S}$, rather than zero. The energy of the quasiparticles is therefore Doppler-shifted from zero, which is readily observed in experiments~\cite{rohlfing_doppler_2009}.

As an example, consider a 2D structure in the $xy$ plane, contacted by a SC from the top (height $L_z$ along the $z$ direction). Applying a magnetic field $B$ along $x$ induces a SC phase gradient along $y$, 
\begin{equation}
    \frac{d\phi}{dy} = -\frac{2\pi}{\Phi_0}BL_{z}.
\end{equation}
Assuming $B$ is uniform (thin SC, no screening), we can then modulate the phase gradient by controlling the height $L_z$ as a function of $x,y$. This way allows one to induce patterns of supercurrents and phase gradients in the system.

\section{Outlook}\label{sec:outlook}

The theoretical proposals we reviewed establish the SC phase difference as a powerful experimental knob for realizing topological superconductivity. By utilizing the simple fact that the SC order parameter is a complex number, various options exist for either reducing the required magnetic field for establishing topological superconductivity,  or eliminating it altogether. The most appealing aspect of this is just how non-invasive phase biasing can be to the parent SC and even to the proximity-induced SC gap. We close this review by pointing out some intriguing future directions related to this field.

A major player that was only briefly touched upon in this review is disorder. Due to the prominence and very advanced experimental status of nanowire-based Majorana devices, there is a massive body of literature on the effect of disorder on them~\cite{brouwer_probability_2011,brouwer_topological_2011,stanescu_majorana_2011,potter_engineering_2011,lutchyn_momentum_2012,sau_experimental_2012,pan_disorder_2021,das_sarma_disorder-induced_2021}. It is generally understood that in nanowires disorder is quite detrimental to the topological phase, and therefore much effort has been devoted to fabricating very clean samples with a good semiconductor-superconductor interface. Moving to the world of phase-induced proposals, the picture is still not entirely clear. A first step towards understanding the role of disorder in such systems was taken by Haim and Stern~\cite{haim_benefits_2019}, who analyzed the phase-biased planar Josephson junction (see Sec.~\ref{subsec:planarJJ} and Refs.~\cite{hell_two-dimensional_2017,pientka_topological_2017}) in the presence of impurity scattering. They examined the dependence of the Majorana localization length on the disorder strength, and found a non-monotonic behavior: when turning on disorder, the localization length first \emph{decreases}, until an ``optimal" point is reached from where the localization length starts growing, eventually surpassing its disorder-free value. The interpretation of this result is that disorder scatters off long trajectories that go back and forth along the junction and do not encounter the SCs. These finite-momentum trajectories have a very small energy gap, which in turn causes the localization length to be large, and therefore eliminating them is beneficial. Too strong disorder, with a typical scattering rate comparable to the induced superconducting gap, eventually suppresses completely the $p$-wave superconductor, in a way which is analogous to the way magnetic impurities suppress convectional superconductivity~\cite{tinkham_introduction_2004}. It remains to be explored whether these findings also holds for other phase-assisted or phase-only proposals.
A hint that phase-only proposals may be more resilient to disorder is the possibility of making the length of the system $L$ small, even smaller than the mean free path, using the rule of thumb $L\Delta\approx\alpha$, such that disorder should be less harmful to the topological state.

So far we have focused on MZMs arising in effectively non-interacting systems (the interactions involved in forming the SC state are well described by mean-field theory). In the presence of electron-electron interactions, a rich variety of additional excitations become available~\cite{wang_topological_2010,ryu_interacting_2012,lu_theory_2012,senthil_symmetry-protected_2015,chiu_classification_2016}. Parafermions, the generalization of MZMs, are of particular interest: they are a fractionalized state of matter and they are predicted to be even more powerful than MZMs for the purpose of quantum computation~\cite{alicea_topological_2016}. The effect of electron-electron interactions on phase-induced Majorana systems is still widely unexplored, and it holds great promise, both for deepening our understanding of interacting topological phases and for practical applications.

Finally, we comment on the possible extensions of 1D phase-induced MZM platforms to two dimensions. The prototypical example of a 2D topological SC is the $p_x+ip_y$ SC~\cite{alicea_new_2012,read_paired_2000}. This is a simple 2D model where the relative phases of the $p$-wave pair potential in the $x$ and $y$ directions differ by $\pi/2$ (in fact, any non-zero phase shift between the two is sufficient, but $\pi/2$ gives the optimal topological protection and it is easy to analyze). The model supports three phases: a topologically trivial phase and two topologically non-trivial phases with an opposite chirality. In these chiral topological phases, a system with open boundary conditions has a chiral Majorana mode at its edge. This 1D Majorana mode is very different from the zero-dimensional MZM appearing at the edges of 1D systems: it disperses and is not stuck at zero energy. The inherent topological nature of these phases is revealed in its full glory by introducing vortices into the system: each vortex binds a MZM at its core, and by moving the vortices around, one can potentially braid the MZMs and observe their non-Abelian nature. The phase-induced platforms for 1D topological SCs reviewed here might offer a fresh approach to advancing the endeavor for 2D topological  superconductivity.

\ack
We are grateful to our collaborators who enriched our understanding of the topic of this review, including Anton Akhmerov, Abhishek Banerjee, Erez Berg, Karsten Flensberg, Leonid Glazman, Charlie Marcus, Andr\'e Melo, Andrew Saydjari, Gal Shavit, Ady Stern, Felix von Oppen, Marie Wesson, Amir Yacoby.
Our research on this topic was supported by the European Union's Horizon 2020 research and innovation programme (Grant Agreement LEGOTOP No. 788715), the DFG (CRC/Transregio 183, EI 519/7-1), ISF Quantum Science and Technology (2074/19), the BSF and NSF (2018643)

\section*{References}
\bibliography{library}

\end{document}